\documentclass[twocolumn,showpacs,preprintnumbers,amsmath,amssymb]{revtex4}

\usepackage{graphicx}
\usepackage{dcolumn}
\usepackage{bm}

\begin{document}


\title{Lennard--Jones and Lattice Models of Driven Fluids}

\author{M. D{\'{\i}}ez--Minguito, P.L. Garrido, and J. Marro}
 \affiliation{Institute `Carlos I' for Theoretical and Computational Physics,\\
and Departamento de Electromagnetismo y F\'{\i}sica de la Materia,\\
Universidad de Granada, E-18071 - Granada, Spain.\\
}%

\date{\today}

\begin{abstract}
We introduce a nonequilibrium off--lattice model for anisotropic phenomena
in fluids. This is a Lennard--Jones generalization of the driven
lattice--gas model in which the particles' spatial coordinates vary
continuously. A comparison between the two models allows us to discuss some
exceptional, hardly realistic features of the original discrete system
---which has been considered a prototype for nonequilibrium anisotropic
phase transitions. We thus help to clarify open issues, and discuss on the
implications of our observations for future investigation of anisotropic
phase transitions.
\end{abstract}

\pacs{05.10.-a, 05.60.-k, 05.70.Fh, 47.11.+j, 47.70.Nd, 51, 64.60.-i}
\maketitle


The concept of \textit{nonequilibrium phase transition} (NPT) \cite%
{Haken,gar,Cross} helps our present understanding of many complex phenomena
including, for instance, the jamming in traffic flow on highways \cite%
{Traffic}, the origin of life \cite{Ferreira}, and the pre--humans
transition to mammals \cite{Treves}. Many studies of NPTs have focused on
lattice systems \cite{Liggett,Zia,Privman,Droz,Marro,Hinrichsen,Odor}. This
is because lattice realizations are simpler than in continuum space, e.g.,
they sometimes allow for exact results and are easier to be implemented in a
computer. Furthermore, a bunch of emerging techniques may now be applied to
lattice systems, including nonequilibrium statistical field theory. A
general amazing result from these studies is that lattice models often
capture the essentials of 
social organisms, epidemics, glasses, electrical circuits, transport,
hydrodynamics, colloids, networks, and markets, for example.

The \textit{driven lattice gas} (DLG) \cite{KLS}, a crude model of \textquotedblleft super ionic
currents\textquotedblright \cite{water}, has become the theoreticians' prototype for anisotropic NPTs. The
DLG consists of a lattice gas with the particles hopping preferentially
along one of the lattice directions, say $\hat{x}.$ One may imagine this is
induced by an external drive, $E\hat{x},$ e.g., an applied electric field
assuming the particles are positive ions. Consequently, for periodic
boundary conditions, a particle current and an anisotropic interface set up
along $\hat{x}$ at low temperature, $T<T_{E}.$ That is, a liquid--like phase
which is striped then coexists with its gas. More specifically, assuming
---for simplicity and concreteness--- the square lattice half filled of
particles, Monte Carlo (MC) simulations show that the function $T_{E}$
monotonically increases with $E$ from the Onsager value $T_{0}=T_{\text{%
Onsager}}=2.269Jk_{B}^{-1}$ to $T_{\infty }\simeq 1.4T_{\text{Onsager}}.$
This limit corresponds to a \textit{nonequilibrium} critical point. As a
matter of fact, it was numerically shown to belong to a universality class
other than the Onsager one, e.g., MC data indicates $\beta \simeq 0.33$
(instead of the Ising value $\beta =1/8$ in two dimensions) for the order
parameter critical exponent \cite{Marro,beta4,beta5,beta6}.

Statistical field theory is a complementary approach to the understanding of
nonequilibrium ordering in the DLG. The derivation of a general mesoscopic
description is still an open issue, however. Two different approaches have
been proposed. The \textit{driven diffusive system} (DDS) \cite%
{Zia,beta2,beta3}, which is a Langevin type of equation aimed at capturing
all the relevant symmetries, predicts that the current will induce a
predominant mean--field behavior and, in particular, $\beta =1/2.$ The 
\textit{anisotropic diffusive system} (ADS) \cite{paco}, which follows after
a non--rigorous coarse graining of the master equation, rules out the
relevance of the current and leads to the above--indicated MC critical
exponent for $E\rightarrow \infty .$ However, the ADS approach reduces to the DDS for
finite $E,$ a fact which is hard to be fitted to MC data, and both contain
disquieting features \cite{problemas}.

In any case, field theoretical studies have constantly demanded further
numerical efforts, and the DLG is nowadays the most thoroughly studied
system showing an anisotropic NPT. The topic is not exhausted, however. On
the contrary, there remain unresolved matters such as the above mentioned
issues concerning critical and mesoscopic behaviors, and the fact that $T_{E}
$ increases with $E$, which is counterintuitive \cite{Zia2}. Another
significative question concerns the observation of triangular anisotropies
at early times after a rapid MC quench from the homogenous state to $T<T_{E}.
$ The triangles happen to point against the field, which is contrary to the
prediction from the DDS continuum equation \cite{triang1}.

\begin{figure}
\includegraphics[width=7.5278cm]{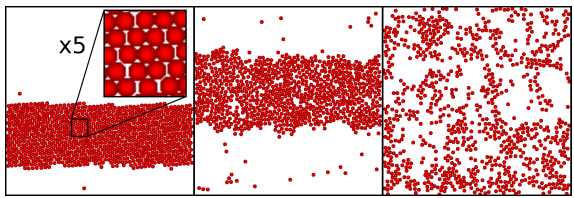}
\caption{\label{fig1} Typical configurations
during the stationary regime of the DLJF subject to a horizontal field of
intensity $E^{\ast }=1.$ These graphs, which are for $N=900$ particles and
density $\protect\rho ^{\ast }=0.30,$ illustrate, from left to right, 
\textit{(i) }coexistence of a solid and its vapor (the configuration shown
is for temperature $T^{\ast }=0.20),$ \textit{(ii) }liquid--vapor
coexistence $(T^{\ast }=0.35),$ and \textit{(iii) }a disordered, fluid phase 
$(T^{\ast }=0.50).$ The left--most graph shows a detail of the solid strip. The particles, which move in a square box of side $\sqrt{N/\rho^{\ast}}$, are given here an arbitrary size, namely, diameter$=1.1\protect\sigma$.}
\end{figure}

This paper reports on a new effort towards better understanding basic features of
NPTs. With this aim, we here present a description of driven systems with
continuous variation of the particles' spatial coordinates ---instead of the
discrete variations in the DLG. We hope this will provide a more realistic
model for computer simulation of anisotropic fluids. Our strategy to set up
the model is to follow as closely as possible the DLG. That is, we analyze
an off--lattice representation of the DLG, namely, a microscopically
continuum with the same symmetries and hopefully criticality. Investigating
these questions happens to clarify the puzzling situation indicated above
concerning the outstanding behavior of the DLG. The clue is that the DLG is,
in a sense, pathological.

\begin{figure}
\includegraphics[width=7.5278cm]{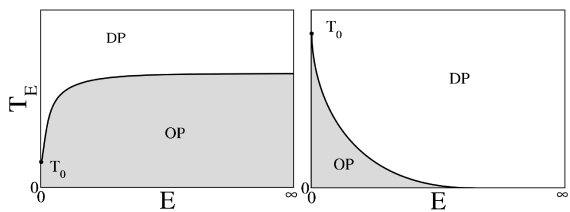}
\caption{\label{fig2} Schematic phase
diagrams for the three models, as defined in the main text, showing ordered
(OP) and disordered (DP) phases. The left graph is for the DLG, for which $%
T_{0}=T_{\text{Onsager}}.$ The graph on the right is valid for both the NDLG,
i.e., the DLG with next--nearest--neighbor (NNN) hops, for which $%
T_{0}=2.35T_{\text{Onsager}},$ and the DLJF, for which $T_{0}^{\ast}=0.459$ \protect\cite{smit}.}
\end{figure}

Consider a \textit{fluid} consisting of $N$ particles in a two--dimensional $%
L\times L$ box with periodic boundary conditions. Interactions are according
to a truncated and shifted Lennard--Jones (LJ) potential \cite{Allen}:%
\begin{equation*}
V(r_{ij})=\left\{ 
\begin{array}{cl}
V_{LJ}(r_{ij})-V_{LJ}(r_{c}), & \text{if }r_{ij}<r_{c} \\ 
0, & \text{if }r_{ij}\geq r_{c}.%
\end{array}%
\right. 
\end{equation*}%
Here, $r_{ij}=\left\vert \vec{r_{i}}-\vec{r_{j}}\right\vert $ is the
relative distance between particles $i$ and $j,$ $V_{LJ}(r)=4\epsilon \left[
(\sigma /r)^{12}-(\sigma /r)^{6}\right] ,$ $\epsilon $ and $\sigma $ are our
energy and length units, respectively, and $r_{c}$ is the \textit{cut-off}
that we shall fix at $r_{c}=2.5\sigma $. The preferential hopping will be
implemented as in the lattice, i.e., by adding a drive to the potential
energy. Consequently, the familiar energy balance is (assuming $k_{B}\equiv 1
$ hereafter):

\begin{equation}
-T^{-1}\left[ \left( H(\vec{c}^{\prime })-H(\vec{c})\right) +E\hat{x}\cdot 
\vec{\delta}\right] .  \label{eq:rate}
\end{equation}%
Where $\vec{c}=\left\{ \vec{r_{1}},...,\vec{r_{N}}\right\} $ stands for configurations, $H(\vec{c})=\sum_{i<j}V(r_{ij})$, and $\vec{\delta}=\vec{r_{i}%
}^{\prime }-\vec{r_{i}}$ is the attempted particle displacement. Defining
the latter is a critical step because, as we shall show, the resulting
(nonequilibrium) steady state will depend, even qualitatively on this
choice. Lacking a lattice, the field is the only source of anisotropy, and
any trial move should only be constrained by a maximum displacement in the
radial direction. That is, we take $0<|\vec{\delta}|<\Delta ,$ where $\Delta
=0.5\sigma $ in the simulations reported here. 
The temperature $T$, number density $\rho=N/L^{2}$, and field variables will be reduced according to $T^{\ast }=T/\epsilon$, $\rho^{\ast
}=\rho\sigma^{2}$, and $E^{\ast}=E\sigma/\epsilon$, respectively. Our model thus reduces for $E\rightarrow 0$ to
the \textit{truncated and shifted LJ fluid}, one of the most studied models
in the computer simulation of fluids \cite{Allen,smit}.

We studied this \textit{driven LJ fluid} (DLJF) in the computer by the MC
method using a \textquotedblleft canonical ensemble\textquotedblright ,
namely, fixed values for $N,$ $\rho^{\ast },$ $T^{\ast },$ and $E^{\ast }.$
Simulations involved up to $N=10^{4}$ particles with parameters ranging as
follows: $0.5\leq E^{\ast }\leq 1.5,$ $0.20\leq \rho ^{\ast }\leq 0.60,$ $%
0.15\leq T^{\ast }\leq 0.55.$ The typical configurations one observes are
illustrated in Fig.~\ref{fig1}. As its equilibrium counterpart, the DLJF
exhibits three different phases (at least): vapor, liquid, and solid (sort
of close--packing phase; see the left--most graph in Fig.~\ref{fig1}). At
intermediate densities and low enough $T^{\ast },$ vapor and a condensed
phase segregate from each other. The condensed droplet (see Fig.~\ref{fig1})
is not near circular as it generally occurs in equilibrium, but strip--like extending along the
field direction. A detailed study of each of these phases will be reported
elsewhere \cite{manolo}; we here focus on more general features.

\begin{figure}
\includegraphics[width=7.6cm]{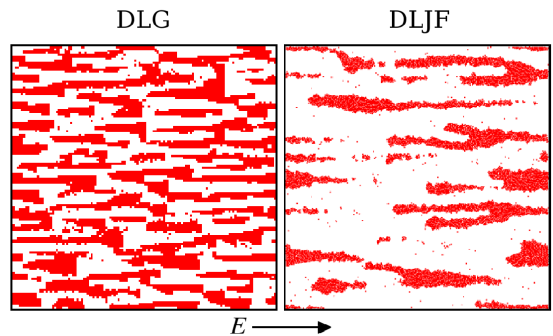}
\caption{\label{fig3} Triangular anisotropies as observed at early times for $E=1$ in computer
simulations of the lattice (left) and the off--lattice (right) models
defined in the main text. The DLG configuration is for $t=6\times 10^{4}$
MCS in a 128$\times $128 lattice with $N=7372$ particles and $T=0.4T_{%
\text{Onsager}}.$ The DLJF configuration is for $t=1.5\times 10^{5}$ MCS, $%
N=10^{4}$ particles, $\protect\rho^{\ast} =0.20$ and $T^{\ast}=0.23.$}
\end{figure}

A main observation is that the DLJF closely resembles the DLG in that both
depict a particle current and the corresponding anisotropic interface.
However, they differ in an essential feature, as illustrated by Fig.~\ref%
{fig2}. That is, contrary to the DLG, for which $T_{E}$ increases with $E,$
the DLJF shows a transition temperature $T_{E}$ which decreases with
increasing $E.$ The latter behavior was expectable. In fact, as $E$ is
increased, the effect of the potential energy in the balance Eq.~(\ref%
{eq:rate}) becomes weaker and, consequently, the cohesive forces between
particles tend to become negligible. Therefore, unlike for the DLG, there is
no phase transition for a large enough field, and $T_{E}\rightarrow 0$ for $%
E\rightarrow \infty $ in the DLJF. Confirming this, typical configurations
in this case are fully homogeneous for any $T$ under a sufficiently large
field $E.$ One may think of variations of the DLJF for which $%
T_{E\rightarrow \infty }=const>0,$ which is more realistic, but the present
one follows more closely the DLG microscopic strategy based on Eq.~(\ref%
{eq:rate}) \cite{manolo}.

Concerning the early process of kinetic ordering, one observes triangular
anisotropies in the DLJF that point along the field direction. That is, the early--time anisotropies in the off--lattice case (right graph in Fig.~\ref{fig3}) are similar to the ones predicted by the DDS, and so they point along the field, contrary to the ones observed in the discrete DLG (left graph in Fig.~\ref{fig3}). 

\begin{figure}
\includegraphics[width=7.5278cm]{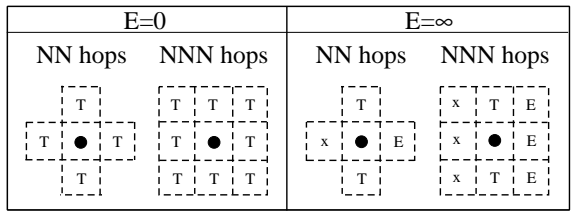}
\caption{\label{fig4} Schematic comparison
of the accessible sites a particle (at the center, marked with a dot) has
for nearest--neighbors (NN) and next--nearest--neighbors (NNN) hops at
equilibrium (left) and in the presence of a large horizontal field (right). The particle--hole exchange between neighbors may be forbidden (x), depend only on the potential energy (T), or occur with probability 1 (E).}
\end{figure}

The above observations altogether suggest a unique exceptionality of the DLG
behavior. This is to be associated with the fact that a driven particle is
geometrically restrained in the DLG. In order to show this, we studied the
lattice with an infinite drive extending the hopping to
next--nearest--neighbors (see also Refs.\cite{triang2,szabo}). As
illustrated in Fig.~\ref{fig4}, this introduces further relevant directions
in the lattice, so that the resulting model, to be named here NDLG, is
expected to behave closer to the DLJF. This is confirmed. For example, one
observes in the discrete NDLG that, as in the continuum DLJF, $T_{E}$
decreases with increasing $E$ ---though from $T_{0}=2.35T_{\text{Onsager}}$
in this case. This is illustrated in Fig.~\ref{fig2}.

\begin{figure}
\includegraphics[width=7.5278cm]{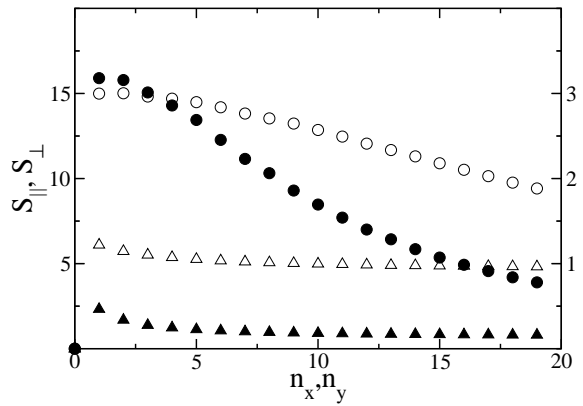}
\caption{\label{fig5} Parallel (circles) and
transverse (triangles) components of the structure factor above criticality
for the DLG on a $128\times 128$ lattice with NN interactions (filled
symbols, left scale) at $T=1.58$ and NNN interactions (empty symbols, right
scale) at $T=1.$ Temperatures normalized to $T_{\text{Onsager}}$. Here, $%
n_{x,y}=128k_{x,y}/2\protect\pi $.}
\end{figure}

There is also interesting information in the two--point correlation function
and its Fourier transform, $S(\vec{k}).$ In fact, the DLG displays (more
clearly above criticality) slow decay of two--point correlations \cite{pedro}
due to the detailed balance violation for $E\neq 0.$ For a half--filled
lattice, this function is $C(\vec{r})=\left\langle s_{\vec{q}}\,s_{\vec{r}+%
\vec{q}}\right\rangle -1/4,$ where $s_{\vec{q}}$ is the occupation number at
site $\vec{q}$ and the steady average $\left\langle ...\right\rangle $
involves also averaging over $\vec{q}.$ Analysis of the components along the
field, $C(x,0),$ and transverse to it, $C(0,y),$ shows that correlations are
qualitatively similar for the DLG and the NDLG ---although somewhat weaker
along the field for NNN hops. That is, allowing for a particle to surpass
its forward neighbor does not modify correlations. The power--law behavior
translates into a discontinuity of $S(\vec{k})$ \cite{pedro}, namely, $%
\lim_{k_{x}\rightarrow 0}S(k_{x},0)\neq \lim_{k_{y}\rightarrow 0}S(0,k_{y}),$
which is clearly confirmed in Fig.~\ref{fig5} for both NN and NNN hops.

The above shows that the nature of correlations is not enough to determine
the phase diagram. There are already indications of this from the study of
equilibrium systems, and also from other nonequilibrium
models. That is, one may have a qualitatively different phase diagram but
essentially the same two--point correlations by modifying the microscopic
dynamics \cite{manolo,szabo}. It follows, in particular, that the
exceptional behavior of the DLG cannot be understood just by invoking the
functions $C(\vec{r})$ and $S(\vec{k})$ or crude arguments concerning
symmetries. The fact that particles are constrained to travel precisely
along the two principal lattice directions in the DLG is the cause for its
singular behavior. Allowing jumping along intermediate directions, as in
both the NDLG and the DLJF, modifies essentially the phase diagram but not
features ---such as power--law correlations--- that seem intrinsic of the
nonequilibrium nature of the phenomenon.

Rutenberg and Yeung \cite{triang2} also performed quenching experiments for
variations of the DLG. They showed, in particular, that minor modifications
in the DLG dynamics may lead to an inversion of the triangular anisotropies
during the formation of clusters which finally condense into strips. Our
observations above prove that such nonuniversal behavior goes beyond
kinetics, namely, it also applies to the stationary state.

The unique exceptionality of the DLG has some important consequences. One is
that this model involves features that are not frequent in nature. There are
situations in which a drive induces stripes but not necessarily DLG
behavior. The fact that a particle impedes the freedom of the one behind to
move along $\hat{x}$ for a large enough field may only occur very seldom in
cooperative transport. The NDLG is more realistic in this sense. In any
case, the great effort devoted to the DLG during two decades has revealed
important properties of both nonequilibrium steady states and anisotropic
phase transitions, and there are some unresolved issues yet. For example, a
general mesoscopic description which captures the exceptionality of the DLG
remains elusive. According to our observations above, such a description
needs to include the microscopic details of transverse dynamics which, in
particular, should allow one to distinguish between the DLG and the NDLG. On
the other hand, it ensues that, due to its uniqueness, the DLG does not have a simple \textit{off-lattice} analog. This is because the ordering agent in the
DLG is more the \underline{lattice} geometry than the field itself. The fact
that one needs to be very careful when modeling nonequilibrium phenomena
---one may induce both a wrong critical behavior and an spurious phase
diagram--- ensues again in this example. This seems not to be so dramatic in
equilibrium where, for example, the lattice gas is a useful
oversimplification of a LJ fluid.


Finally, we remark that a novel fluid model in which the particles move in a
continuum space has been introduced in this paper. The particle `infinite
freedom' is realistic, as it is also the LJ potential. It may contain some
of the essential physics in a class of nonequilibrium anisotropic phenomena and phase transitions. On the other hand, the model is simple enough
to be useful in computer simulations, and it is endowed of even--simpler and
functional lattice analogs such as the NDLG.

We acknowledge very useful discussions with E. Albano, M. A. Mu\~noz,
and F. de los Santos, and financial support from MEyC and FEDER (project FIS2005-00791).


\begin{thebibliography}{99}
\bibitem{Haken} H. Haken, Rev. Mod. Phys. \textbf{47}, 67 (1975).

\bibitem{gar} L. Garrido, \textit{Far from Equilibrium Phase Transitions},
Sitges Conference 1988 (Springer Verlag, Berlin, Germany, 1989).

\bibitem{Cross} M.C. Cross and P.C. Hohenberg, Rev. Mod. Phys. \textbf{65},
851 (1993).

\bibitem{Traffic} K. Nagel and M. Schreckenberg, J. Phys. I (France) \textbf{%
2}, 2221 (1992); D. Chowdhury \textit{et al.}, Phys. Rep \textbf{329}, 199
(2000).

\bibitem{Ferreira} C. P. Ferreira and J. F. Fontanari, Phys. Rev. E \textbf{%
65}, 021902 (2002).

\bibitem{Treves} A. Treves and Y. Roudi, in \textit{Methods and Models in
Neurophysics}, Les Houches School 2003 (Elsevier, 2005).

\bibitem{Liggett} T. M. Liggett, \textit{Interacting Particle Systems},
(Springer Verlag, Heidelberg, Germany, 1985).

\bibitem{Zia} B. Schmittmann and R. K. P. Zia, in \textit{Statistical
Mechanics of Driven Diffusive Systems} in Phase Transitions and Critical
Phenomena, Vol. 17, edited by C. Domb and J. L. Lebowitz (Academic, London,
U.K., 1996).

\bibitem{Privman} V. Privman, \textit{Nonequilibrium Statistical Mechanics
in One Dimension}, (Cambridge University Press, Cambridge, U.K., 1996).

\bibitem{Droz} B. Chopard and M. Droz, \textit{Cellular Automata Modeling of
Physical Systems}, (Cambridge University Press, Cambridge, U.K., 1998).

\bibitem{Marro} J. Marro and R. Dickman, \textit{Nonequilibrium Phase
Transitions in Lattice Models}, (Cambridge University Press, Cambridge,
U.K., 1999).

\bibitem{Hinrichsen} H. Hinrichsen, Adv. Phys. \textbf{49}, 815 (2000).

\bibitem{Odor} G. \'{O}dor, Rev. Mod. Phys. \textbf{76}, 663 (2004).

\bibitem{KLS} S. Katz, J. L. Lebowitz, and H. Spohn, J. Stat. Phys. \textbf{34%
}, 497 (1984).

\bibitem{water} A. F. Goncharov \textit{et al}., Phys. Rev. Lett. \textbf{94},
125508 (2005).

\bibitem{beta4} A. Achahbar, P. L. Garrido, J. Marro, and M. A. Mu\~{n}oz,
Phys. Rev. Lett. \textbf{87}, 195702 (2001).

\bibitem{beta5} E. V. Albano and G. Saracco, Phys. Rev. Lett. \textbf{88},
145701 (2002).

\bibitem{beta6} E. V. Albano and G. Saracco, Phys. Rev. Lett. \textbf{92},
029602 (2004).

\bibitem{beta2} K. Leung and J. L. Cardy, J. Stat. Phys. \textbf{44}, 567
(1986).

\bibitem{beta3} H. K. Janssen and B. Schmittmann, Z. Phys. B \textbf{63},
517; \textit{ibid} \textbf{64}, 503 (1986).

\bibitem{paco} F. de los Santos, P. L. Garrido, and M.A. Mu\~{n}oz, Physica
A \textbf{296}, 364 (2001).

\bibitem{problemas} A. Achahbar \textit{et al.}, to be published.

\bibitem{Zia2} B. Schmittmann and R. K. P. Zia, Phys. Rep. \textbf{301}, 45
(1998).

\bibitem{triang1} F. J. Alexander, C. A. Laberge, J. L. Lebowitz, and R. K.
P. Zia, J. Stat. Phys. \textbf{82}, 1133 (1996).

\bibitem{Allen} M. P. Allen and D. J. Tidlesley, \textit{Computer
Simulations of Liquids}, (Oxford University Press, Oxford, U.K., 1987).

\bibitem{smit} B. Smit and D. Frenkel, J. Chem. Phys. \textbf{94}, 5663
(1991).

\bibitem{manolo} M. D\'{\i}ez--Minguito \textit{et al.}, to be published.

\bibitem{triang2} A. D. Rutenberg and C. Yeung, Phys. Rev. E \textbf{60},
2710 (1999).

\bibitem{szabo} A. Szolnoki and G. Szab\'{o}, Phys. Rev. E \textbf{65},
047101 (2002).

\bibitem{pedro} P. L. Garrido, J. L. Lebowitz, C. Maes, and H. Spohn, Phys.
Rev. A \textbf{42}, 1954 (1990).

\end{thebibliography}
\end{document}